# Terrestrial Planet Finder – Coronagraph (TPF-C) Flight Baseline Mission Concept

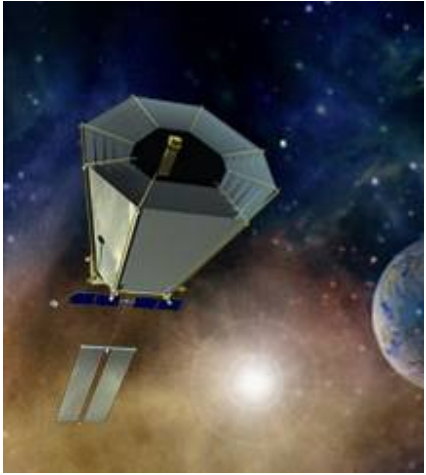

## Marie Levine
(818) 354-9196
*Marie.Levine@jpl.nasa.gov*

## Doug Lisman
## Stuart Shaklan

*Jet Propulsion Laboratory,*
*California Institute of technology*

April 1, 2009

## Co-Authors

**TPF-C Science and Technology Definition Team (STDT)**

Roger Angel, University of Arizona
Michael Brown, California Institute of Technology
Robert Brown, Space Telescope Science Institute
Christopher Burrows, Metajiva
Mark Clampin, NASA Goddard Space Flight Center
Henry Ferguson, Space Telescope Science Institute
Olivier Guyon, University of Arizona
Heidi Hammel, Space Science Institute
Sara Heap, NASA Goddard Space Flight Center
Scott Horner, Lockheed-Martin
N. Jeremy Kasdin, Princeton University
**James Kasting, Penn State University (Chair)**
Mark Kuchner, NASA Goddard Space Flight Center
Douglas Lin, University of California, Santa Cruz
Mark Marley, NASA Ames Research Center
Victoria Meadows, University of Washington
Martin C. Noecker, Ball Aerospace & Technology Corp.
Ben Oppenheimer, American Museum of Natural History
Sara Seager, Massachussetts Institute of Technology
Michael Shao, Jet Propulsion Laboratory
Karl Stapelfeldt, NASA Jet Propulsion Laboratory
John. Trauger, NASA Jet Propulsion Laboratory
**Wesley Traub, NASA Jet Propulsion Laboratory (Co-Chair)**

**TPF-C Flight Baseline #1 Design Team & Collaborators**

James Alexander, NASA Jet Propulsion Laboratory
Carl Blaurock, Nightsky Systems
Terry Cafferty, TC Technology
Eri Cohen, NASA Jet Propulsion Laboratory
David Content, NASA Goddard Space Flight Center
Larry Dewell, Lockheed Martin
Philip Dumont, NASA Jet Propulsion Laboratory
Robert Egerman, ITT Space Systems, LLC
Virginia Ford, TMT Observatory Corporation
Joseph Greene, NASA Jet Propulsion Laboratory
Timothy Ho, NASA Jet Propulsion Laboratory
Sarah Hunyadi, NASA Jet Propulsion Laboratory
Sandra Irish, NASA Goddard Space Flight Center
Clifton Jackson, NASA Goddard Space Flight Center
Andrew Kissil, NASA Jet Propulsion Laboratory
Michael Krim
Eug-Yun Kwack, NASA Jet Propulsion Laboratory
Chuck Lillie, Northrop Grumman Aerospace Systems
Alice Liu, NASA Goddard Space Flight Center
Luis Marchen, NASA Jet Propulsion Laboratory
Gary Mosier, NASA Goddard Space Flight Center
Pantazis Mouroulis, NASA Jet Propulsion Laboratory
Raymond Ohl, NASA Goddard Space Flight Center
Joe Pitman, Exploration Sciences
Stephen Ridgway, National Optical Astronomy Observatories
Erin Sabatke, Ball Aerospace
Andrew Smith, NASA Goddard Space Flight Center
Remi Soummer, Space Telescope Science Institute
Domenick Tenerelli, Lockheed Martin
Robert Vanderbei, Princeton University

## 1. EXECUTIVE SUMMARY

The Terrestrial Planet Finder Coronagraph *(*TPF-C) mission presented here is an existence proof for a flagship-class internal coronagraph space mission capable of detecting and characterizing Earth-like planets and planetary systems at visible wavelengths around nearby stars, using an existing launch vehicle. TPF-C will use spectroscopy to measure key properties of exoplanets including the presence of atmospheric water or oxygen, powerful signatures in the search for habitable worlds.

Following the strong endorsement from the last decadal survey on *Astronomy and Astrophysics for the New Millennium* (2001), NASA pursued a vigorous TPF program with nearly $150M community investment in technology, science and mission studies over the last decade. In 2004, after a 3-year head-to-head comparison of various coronagraph, interferometer and occulter architectures NASA chose a visible imaging coronagraph as the first of its exoplanet characterization missions [1, 2]. The ensuing study evaluated the design and technology of a band-limited (BL) Lyot coronagraph operating at an inner working angle (IWA) of $4\lambda/D$ over 0.5-1.1 μm. A BL coronagraph was selected for the TPF-C Flight Baseline #1 (FB1) architecture as it was the most mature technique, having been demonstrated in the laboratory to perform at levels needed for detecting Earths, $5.2 \times 10^{-10}$ at $4\lambda/D$ for 760-840nm (10% band) in natural unpolarized light, thus verifying the fundamental physics and establishing its feasibility [8, 11]. In 2006, the TPF-C Science and Technology Definition Team (STDT) established the science requirements and reviewed the mission concept [3], summarized here in §2. TPF-C mission activities were abruptly terminated by NASA in 2006, with minimal support to on-going technology efforts. In 2008, the Exoplanet Task Force (ExoPTF) recommended immediate investment in coronagraph technology towards a direct detection space mission, such as TPF-C, beginning in the next decade [10].

TPF-C FB1 implements the largest possible 8m x 3.5m monolithic off-axis primary mirror with multiple flight system deployments that fits within the nation's biggest existing launch shroud [4]. In addition to the coronagraph, the TPF-C instrument suite also includes a spectrometer and a general astrophysics wide-field camera. Engineering analyses found that FB1 met the contrast and stability requirements needed to study Earths. A detailed Integration and Test (I&T) plan defined a verification approach using existing facilities, whereby the coronagraph instrument is tested full-scale and the integrated observatory relies heavily on subscale test articles anchoring models that scale performance from ground to space (§3). Overall no engineering show stoppers were found.

To date, TPF-C has distributed nearly half of its total technology budget to the nation's coronagraph community, maturing internal coronagraph instrument options other than the BL used in FB1, especially those that promise to provide higher efficiency at smaller IWA. A technology mirror demonstrator was also initiated. TPF-C established technology infrastructure applicable to many other coronagraphs, now enabling the demonstration of the NASA Astrophysics Strategic Mission Concept Studies (ASMCS). In particular, the High Contrast Imaging Tested (HCIT) is the premier facility for demonstrating starlight suppression to Earth-twin performance levels. Modeling tools exist for rapid prototyping of coronagraph contrast errors budgets and for seamless integrated opto-thermo-structural-controls analyses within a single multi-physics observatory system model. Coronagraph models are being validated against test data showing better than the needed $10^{-9}$ contrast. A detailed TPF-C technology plan, approved by NASA HQ in 2005, traces the maturation approach of the key technologies to TRL 6 by the end of Phase B [5]. The technology needs for visible coronagraphs are well understood, where the top three for FB1 are starlight suppression (SS), precision system modeling, and large space optics (§4). A coronagraph operating at $2\lambda/D$, instead of $4\lambda/D$, allows almost the same exoplanet science with a telescope half the size of FB1. Preliminary work began in 2006 on a second TPF-C design cycle with a circular off-axis 4m-class telescope and a coronagraph at $2.5\lambda/D$ which avoids any telescope deployments yet fits inside an existing Atlas V shroud, promising to reduce cost and risk, albeit with tighter stability requirements.

We appeal to the Astro2010 to support a focused exoplanet coronagraph technology development and mission definition program, leading to a TPF-C launch at the earliest opportunity.

## 2. KEY SCIENCE GOALS

The science case for TPF-C was established by the TPF-C Science and Technology Definition Team (STDT) and is reported in detail in the STDT Report [3].

**The scientific goals of the TPF-C mission—to discover and study Earth-sized planets around neighboring stars—are ambitious, exciting and profound, addressing some of the most important questions humankind can ask about its place in the universe.** Scientists have found a variety of giant planets, and are poised to find smaller planets, more and more like the Earth. TPF-C will be our first chance to detect large numbers of Earth-sized planets nearby, see them directly, measure their colors, study their atmospheres, and look for evidence of life there. These goals make TPF-C a special project in the history of astronomy, one capable of firing human imagination and revolutionizing the way we think about ourselves and the universe.

The existence of planets around other stars, an unsupported scientific hypothesis until the mid-1990s, is no longer in doubt. Nearly 350 extrasolar planets have been discovered around other main sequence stars, most of these using the ground-based radial velocity (RV or Doppler) technique. The next frontier for planet-finding is to look for rocky, terrestrial-type planets around other stars. NASA's Kepler mission and ESA's Corot mission will do this for more than a hundred thousand very distant stars, while the Space Interferometry Mission (SIM PlanetQuest) searches around nearby stars. Both Kepler and *SIM* have the capability to detect at least a few Earth-size planets if they are common. Ongoing ground-based searches may also reveal Earth-mass planets around very low-mass stars.

How well TPF-C will be able to characterize the planets it discovers depends on the design of both the telescope and the spectrograph. The baseline design has a wavelength range of 0.5-1.1 µm and a spectral resolving power, $\lambda/\Delta\lambda$, of 70. For an Earth twin (planet and star exactly like our Earth and Sun) seen at 10 pc distance, these capabilities would enable TPF-C to measure absorption bands of water vapor, oxygen, and possibly ozone. The presence of water vapor is an indicator of potential habitability, as liquid water is considered to be a prerequisite for life as we know it. Oxygen and ozone are potential indicators of life itself, because on Earth they come mainly from photosynthesis. There may be planets on which $O_2$ and $O_3$ can build up abiotically, but for most planets within the liquid water habitable zone, these gases are considered to be reliable bioindicators. Hence, TPF-C has the potential to provide compelling evidence of life on extrasolar planets, answering this age-old question that encompasses science, philosophy, and issues of human identity.

**TPF-C can also study giant planets and dust disks — the entire *planetary system architecture* — at the same time that it looks for Earth-like planets, supporting our studies of the potential habitability of any Earth-like planet.** If our own Solar System is a guide (it still is, by what we know today), planets like Earth are found in planetary systems that include other small rocky planets, e.g., Venus and Mars, along with gas giants like Jupiter and Saturn, and ice giants like Uranus and Neptune. The larger planets are of interest in their own right, but they may also be crucially connected to the habitability of the Earth-like planets. In our own Solar System, for example, Jupiter helps shield Earth from collisions with comets, but also perturbs some asteroids into Earth-crossing orbits. Thus understanding the potential habitability of an Earth-like planet requires study of the entire planetary system architecture. Fortunately, these studies can be done at the same time as terrestrial planet-finding observations that they support.

**TPF-C will also study the dust clouds around stars, to learn about the process of planetary formation.** Planetary systems themselves do not occur in isolation around stars. Collisions between small bodies (asteroids) within the system and vaporization of icy planetesimals (comets) from farther out create dust that orbits the star along with the planets. This dust reflects starlight, giving rise to the zodiacal light in our own Solar System and to exozodiacal light in other planetary systems. The planets in a given system must be observed against these backgrounds of the "zodi" and the "exozodi." The exozodiacal light in a given system must be measured and "removed" in order to see the planets. However, it is also known that the dust distribution can be perturbed by the gravitational influence of planets; thus the exozodi light may be a powerful tool for finding and studying the planets in a system. For these reasons, the study of exozodiacal dust clouds is an integral part of the TPF-C mission. Mapping out the exozodiacal light can be carried out simultaneously with the search for terrestrial planets.

**In addition to its primary goal of searching for terrestrial planets and the dusty systems that accompany them, TPF-C will make substantial contributions in other areas of general astrophysics.** The telescope will be very large, smooth, and stable, and so will exceed the performance of Hubble Space Telescope (HST) in several respects, including collecting area, angular resolution, and PSF stability. To fully take advantage of this large telescope, a separate instrument—a wide-field camera—is planned, in addition to the coronagraph. This instrument would channel light along a different optical path, and hence could perform its tasks either in parallel with planet-finding activities or by using the telescope in pointed mode. An example of a parallel general astrophysics science observation is the imaging of distant galaxies, similar to the Hubble Deep Fields but with even greater depth and clarity. Such deep fields could be obtained during the extended time intervals, one day to several weeks, required for planetary detection and characterization.

The overall science objectives for TPF-C are summarized in Table 1. The Science requirements and derived engineering requirements are summarized in Table 2. Our driving requirement is to sample 30 habitable zones in a 3-year mission assuming that 1/3 of the time is used for planet detection. This requires $\Delta$mag = 25.5, IWA =4 $\lambda$/D for lambda = 600 nm, and a bandwidth of 100 nm. These values can all be traded against each other, and they are a function of the mask and system throughput, assumed exozodiacal level, and other parameters.

**Table 1 Summary of TPF-C Science Objectives**

| Science | # | Objective |
|---|---|---|
| Terrestrial Plant Science | 1 | Directly detect terrestrial planets within the habitable zones around nearby stars or, show they are not present. |
| | 2 | Measure orbital parameters and brightnesses for any terrestrial planets that are discovered. |
| | 3 | Distinguish among planets, and between planets and other objects, through measurements of planet color. |
| | 4 | Characterize at least some terrestrial planets spectroscopically, for O2, O3, H2O, and possibly CO2 & CH4. |
| Giant Planets & Planetary System Architecture | 5 | Directly detect giant planets of Jupiter's size and albedo at a minimum of 5 AU around solar type stars, and determine orbits for such giant planets when possible |
| | 6 | Obtain photometry for the majority of detected giant planets, to an accuracy of 10% in at least three broad spectral bands, and in additional bands for the brightest or well-placed giants. |
| | 7 | Characterize detected giant planets spectroscopically, searching for the absorption features of CH4 and H2O. |
| Disk Science and Planet Formation | 8 | Measure the location, density, and extent of dust particles around nearby stars for the purpose of comparing to, and understanding, the asteroid and Kuiper belts in the Solar System. |
| | 9 | Characterize disk-planet interactions with the goal of understanding how substructures within dusty debris disks & infer the presence of planets. |
| | 10 | Study the time evolution of circumstellar disks, from early protoplanetary stages through mature main sequence debris disks. |
| General Astrophysics (examples) | 11 | Constrain the nature of Dark Energy via precise measurements of the Hubble constant and the angular-diameter vs. redshift relation. |
| | 12 | Use the fossil record of ancient stars in the Milky Way and nearby galaxies to measure the time between the Big Bang and the first major episodes of star formation. |
| | 13 | Determine what sources of energy reionized the universe and study how galaxies form within dark-matter halos, through a program of low-resolution spectroscopy of large statistical samples, gathered in parallel with the *TPF-C* planet search program. |
| | 14 | Carry out a diverse General-Observer program in the tradition of the *Hubble, Chandra, Spitzer,* & *JWST* observatories. |

## 2.1 DESIGN REFERENCE MISSION

Early work on the Design Reference Mission (DRM) for TPF-C provided benchmarks of scientific output, to be used for comparing different observatory designs. A key figure of merit was completeness, defined as the fraction of all possible habitable zone (HZ) orbits that are examined for the presence of a planet at least once during the mission. We can also say the probability of a false negative result for a given star (a planet does exist in the HZ but is never found) is one minus the completeness or that completeness is the expected number of Earths detected for $\eta_{earth} = 1$. Multiplying the mission completeness by $\eta_{earth}$ gives the expected number of Earths to be found. Mission models chiefly focused on the integration time needed to reach a given planet sensitivity for each star in the catalog, the number of visits needed to achieve a certain completeness, and how many stars can be scrutinized at that level. One could compare different telescope and coronagraph architectures based on how many stars they could examine.

**Table 2 Top-level TPF-C FB1 Engineering Requirements**

| Science Requirement | Performance Requirement | Design |
|---|---|---|
| Detect Earth twin at 10 parsec | IWA ≤ 100 mas | Search 32 composite Habitable Zones<br>3.5m x 8m aperture at 4 λ/D<br>Starlight suppression with 8th order Band Limited mask<br>Roll angles sweep narrow-beam axis<br>Roll angle dithers remove instrument noise |
| Detect ≥ 1 HZ planet with 95% confidence, if 10% of target stars have planets | Contrast ≥ 25 mags<br>IWA ≤ 65 | |
| Detect Jupiter twin at 10 pc | OWA ≥ 500 mas | R-C telescope: 500nm diff. limit 1" off-axis<br>96 x 96 element Deformable Mirror |
| Measure planet brightness within 10% | Contrast stability ≥ 28 mags | 1 pm wavefront stability per Zernicke mode<br>1 mK stability (PM & coronagraph)<br>V-groove sunshade for solar isolation |
| Detect atmospheric $O_2$ and $H_2O$ | Bandpass = 0.5 to 1.0 µm<br>Spectral resolution ≥ 70 | 2 coronagraph channels feed 2 IFUs |
| Zodi light limited imaging over bandpass | FOV ≥ 10 arc-min$^2$<br>Diffraction limited to 10' off-axis | Reoptimize, probably with TMA telescope |

These early studies indicated that for TPF-C between 35 and 50 stars could be searched for planets, using 2 years out of a 5-year mission lifetime. A total of 1 year was reserved for characterization (mainly spectra) concurrent with the search phase and 2 years for general astrophysics. We have made substantial progress in understanding how the final scientific output depends on observatory requirements and mission scheduling. Now we understand that *TPF-C*, more than any previous space astronomy mission, relies on just-in-time contingent scheduling of observations to achieve its greatest scientific harvest. This is mainly because many of the exoplanets we find will be a surprise, and then their orbital motions limit the times when we can see them again. This "recovery" of the planet after its first disappearance plays an important role in later chances for more in depth characterization. Our Monte Carlo studies show that when we detect a planet for the first time, we must be quick to schedule a follow-up observation to:

- Differentiate between planets and background confusion sources
- Perform low-resolution color measurements, to categorize the type of planet
- Determine or constrain orbits, to help characterize their habitability.
- Carry-out higher-resolution spectroscopy, to search for atmospheric signatures

Some additional work has been focused on how to "front-load" the observing schedule with stars which are known to harbor planets, either from SIM astrometric detections of small rocky planets or from planet detections made through ground observations, Corot or Kepler. This is a very useful additional guideline in the design of a mission observing schedule.

### 2.2 OBSERVING SCENARIO

To distinguish the speckles induced by observatory imperfections from true planets, we compare two images of the planetary system taken with a different "roll" orientation—rotation around the optical axis (Table 3, Figure 1). This pair of images is called an observation. Through the roll maneuver, speckles are expected to stay fixed on the focal plane, while the planet stays fixed on the sky and moves on the focal plane, rotating around the star image. During an observation, the speckle brightness pattern must be stable to much less than the expected planet brightness, so that the image subtraction will unambiguously distinguish speckles from planets. This technique is called angular differential imaging and is illustrated in [11] where the laboratory environmental stability led to a reduction of the background speckle intensity by a factor of about 50.

**Table 3 Observational Scenario Steps**

| | |
|---|---|
| 1 | Acquire target star |
| 2 | Stabilize dynamics and collect light |
| 3 | Using coarse and fine WF control, suppress starlight |
| 4 | Dither 30 degrees w/o changing WF control positions |
| 5 | Stabilize dynamics and collect light |
| 6 | Subtract images |
| 7 | Roll to next 60 degree orientation |
| 8 | Repeat steps 2 through 6 two times |

### 2.3 MISSION SCALING

We compare the Earth-size planet detection capability of various exoplanet mission options as a function of mission size and inner working angle for a limiting delta magnitude sensitivity of 25.5 and 3-year mission duration [6]. The results are summarized in Figure 2 as a function of telescope size (8m, 4m and 2.5m) for two coronagraph options: An eighth order

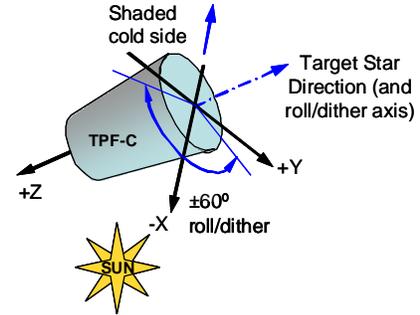

**Figure 1 Observing Scenario**

BL masks ("BL8") as in FB1, a more aggressive Phase Induced Amplitude Apodization (PIAA) operating at 2.5λ/D ("PIAA @ 2.5λ/D") or a coronagraph operating with throughput similar to that of the PIAA at 4λ/D. This later option is not the desired implementation of the PIAA itself and is only viable if the pointing technology required for the more aggressive IWA is not successful. Also included for comparison are the completeness of the TPF Interferometer (TPF-I) with 5 spacecrafts ("classic") or 4 spacecrafts ("Emma"), as well as a 50m external occulter implementation with a 4m telescope and the Space Interferometry Mission (SIM). Note the performance of the 4m mid-scale PIAA option at 2.5 λ/D compared to that of the flagship 8m telescope BL8 at 4λ/D (i.e., TPF-C FB1) and that of the 4m telescope plus a 50m occulter.

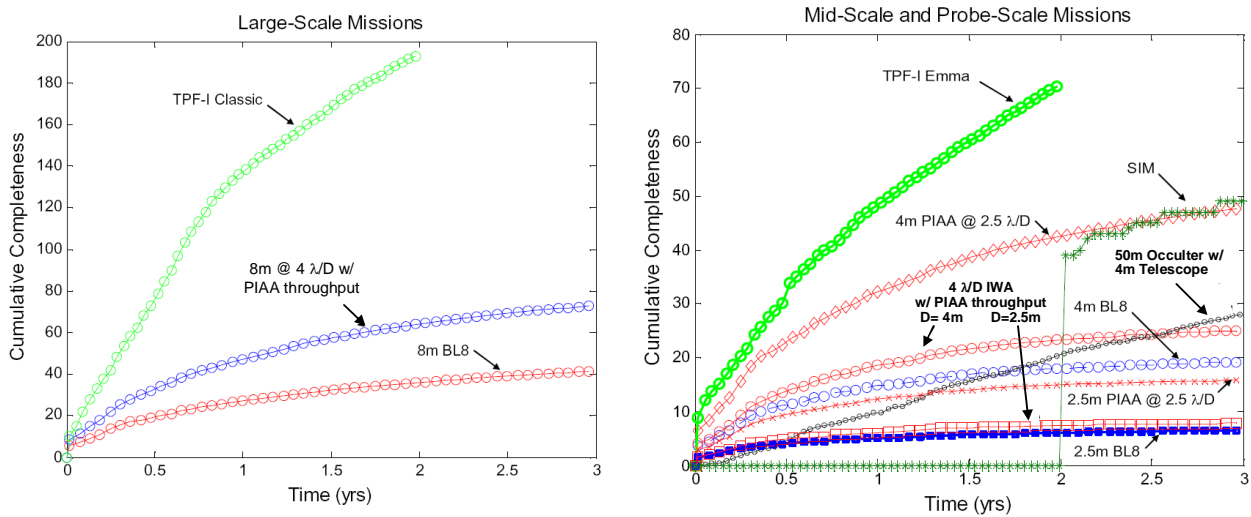

**Figure 2 Comparison of the cumulative completeness for Earth-like planets for various exoplanet mission concepts over a 3-year mission span (Hunyadi, 2007).**

## 3. TPF-C BASELINE OBSERVATORY TECHNICAL OVERVIEW

### 3.1 SYSTEM DESIGN

**The TPF-C design is documented in great detail in the TPF-C STDT Report [3] and the Flight Baseline 1 (FB1) design report [4], both available online.** FB1 refers to a first design cycle, intended to establish a proof of concept, while penetrating the complexities of precision coronagraphy and developing and advancing the requisite modeling tools and methodologies.

TPF-C FB1 is designed to operate at visible wavelengths from 0.5 -1.1 μm with an effective IWA of 65.5 mas or 4λ/D, an outer working angle (OWA) of 500 mas, a scattered light level equal to $6 \times 10^{-11}$ of the stellar peak brightness ($\Delta mag_s = 25.5$), and stability or knowledge of that scattered light to about 10% ($\Delta mag_s = 28$). The coronagraph is chosen to be an 8th order BL mask coronagraph, as the most mature available sensing option. To accommodate present launch vehicle shrouds, an $8 \times 3.5$ m *elliptical* monolithic off-axis shape was adopted for the primary mirror. In general, segmented and/or on-axis primary mirrors are not an option for internal coronagraphs because of edge diffraction effects. To emulate performance of a filled 8m circular aperture, FB1 assumed exposures taken at 3 different roll angles for searching the habitable zones and outer planets of nearby stars. FB1 also includes a wide field camera with a 6 x 10 arcmin field of view that operates either in parallel with the coronagraph or in dedicated mode.

TPF-C operates in a halo orbit at Earth-Sun L2 for a 5-year prime mission, with a design goal of operating for 10 years. Launch is from Cape Canaveral on a Delta-IVH launch vehicle with an existing 5 m fairing. All communications and navigational tracking are via S-Band and near-Earth Ka-Band channels of the 34m DSN.

TPF-C combines an integrated thermal and vibration control system to provide the stable disturbance environment needed for such precise coronagraphy. Telescope stability is accomplished with large deployable concentric conic-shaped films that shed the solar heat input and isolate the payload from changing sun angles during observational maneuvers (A cylindrical version of the v-groove sunshade being implemented on JWST). Coronagraph instrument thermal stability is achieved with a thermal enclosure around the payload that actively controls temperatures in the back end of the telescope. Jitter stability is provided through a two-stage passive isolation system which offers the required vibration reduction from the reaction wheel disturbances. Alternatively, a pointing control system architecture including active dynamic isolation is being considered. An active secondary mirror positioning system aligns the telescope after launch.

The observatory relies on a precision instrument optical control system to provide a stable, high-quality wavefront to the coronagraph. The starlight suppression system (SSS) is a stellar coronagraph designed to eliminate diffracted light and control scattered light, in order to reduce the background light in the instrument to a level that is less than $10^{-10}$ of the incident light. The scattered light is controlled using a coarse deformable mirror (DM) and a pair of fine DMs. The coarse DM compensates for large wavefront deviations left in the telescope, such as due to gravity release and launch stresses. The fine DMs have a 1-micron stroke and high actuator density. As a pair they can control both amplitude and phase wavefront distortions up to a spatial frequency limit determined by the actuator density.

The observatory mass and power estimates were determined both from in depth mechanical modeling and from analogy to previous missions (Tables 4 & 5). The FB1 solar arrays are sized to provide 3000 W of end of life power.

**Table 4 TPF-C FB1 Optimized**

| Component | Mass (kg) | Mass % |
|---|---|---|
| Optical Telescope Assembly | 2400 | 38% |
| Science Payload | 1700 | 27% |
| Spacecraft | 1700 | 27% |
| Total Wet Mass | 5800 | |
| Total Launch Mass | 6400 | |
| Launch Vehicle Capability | | 9200 |
| Launch Margin | | 30% |

**Table 5 TPF-C FB1 Nominal Power**

| Component | Power (W) | Power % |
|---|---|---|
| Telescope Electronics | 80 | 4% |
| Science Payload | 390 | 19% |
| Thermal Control | 580 | 28% |
| Spacecraft | 1000 | 49% |
| Total Observatory Power | 2055 | |
| Available EOL Power | | 3000 |
| Power Margin | | 32% |

Figure 3 and Figure 4 show the FB1 deployed and stowed configuration, with sub-systems.

**Telescope:** The telescope is a 12 m tall assembly, featuring an EFL=140 m optical system with an 8x3.5m elliptical aperture primary mirror (PM) whose focal length is 13.38 m (Figure 5). Trade studies were performed among several telescope optical designs meeting requirements as well as packaging allocations. These include three-mirror anastigmats, Gregorian, Cassegrain, and the baseline Ritchey-Chretien (R-C) design. Considerations of aberration control and packaging led the final choice of the off-axis R-C telescope, but there is room for further optimization with additional study. The telescope is off-axis to avoid obscurations from the secondary mirror (SM) supports, as required for the level of starlight suppression necessary. The PM and SM are coated with protected silver to maximize throughput between 500-1100nm.

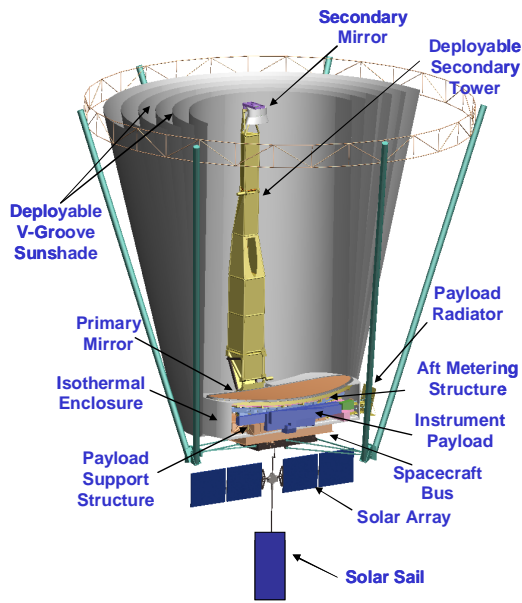

**Figure 3  TPF-C FB1 Deployed Configuration**

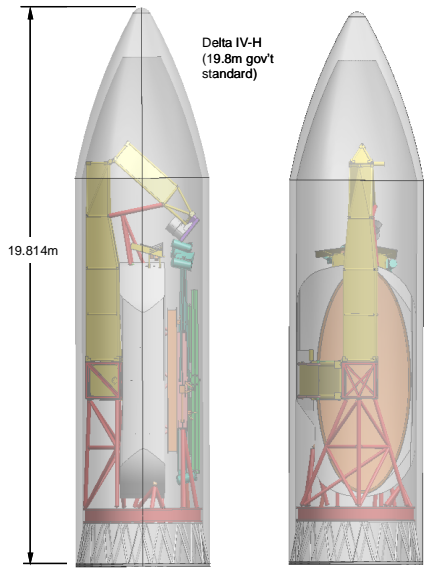

**Figure 4   TPF-C FB1 Stowed Configuration in Delta IV-H Shroud**

The PM is a closed-back design constructed of Corning ULE® glass with a segmented light-weighted honeycomb core that is joint to the front (optical) and back facesheets using ITT/Corning proprietary Low Temperature Fusion process (LTF). Prior to LTF, the facesheets are pocket milled to reduce mass while maintaining sufficient local stiffness over the unsupported regions of the front facesheet to minimized mid-spatial frequency errors from polishing and the uncertainty in the gravity release. FB1 assumed that the PM blank would initially made using flat components, and then Low Temperature Slumped (LTS) over a mandrel to get the required initial curvature. The combined LTF/LTS process has been used on light-weight, relatively compliant active optics such as the Advanced Mirror System Demonstrator, and one of the purposes the TPF funded Technology Demonstration Mirror (TDM) was to demonstrate the viability of this process for a mirror with a stiffness comparable to the TPF-C PM. Unfortunately, due to budget cuts, this program was not completed. Without a demonstration, parts would need to be curved prior to LTF (this has been demonstrated on other programs). Ion beam polishing is used to achieve the wavefront specification. All primary mirror processes have heritage from smaller mirrors and will be demonstrated on a sub-scale prototype.

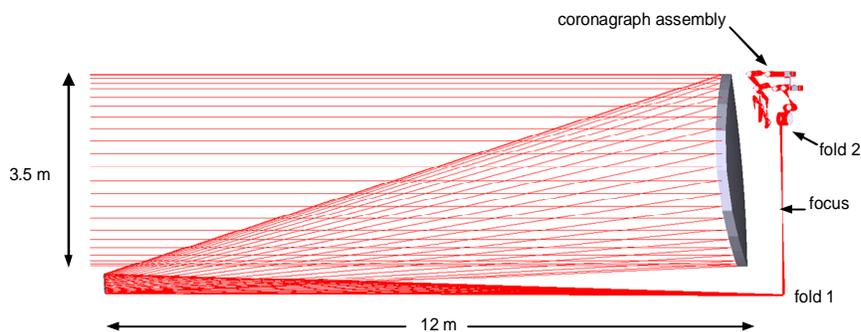

**Figure 5 TPF-C FB1 Telescope Design**

The passive PM is kinematically mounted on 3 flexured bipods extending from the backside of the mirror attached to a strong-back support structure called the Aft Metering Structure (AMS). Models predict gravity effects imparted to the PM of several hundred μm with an uncertainty of +/- 10 μm. Thus the design includes a coarse DM located between the coronagraph and telescope to correct for the on-orbit gravity release uncertainty. The SM assembly is attached atop a folding tower on thermal isolators. The SM is mounted on an active hexapod with 1 Hz bandwidth to correct for gravity release uncertainties and to maintain the long-term alignment of the telescope.

The SM tower folds along 3 hinge lines to stow for launch. Each hinge will be locked out after the tower deploys as the locking mechanisms will then join the main structure together. Behind the secondary mirror is a fine positioning actuated hexapod. The tower assembly attaches to a bracket that kinematically interfaces to the AMS through 3 thermal isolating bipods. Both the PM and SM are enclosed in separate thermal enclosures. The AMS is kinematically attached to the Payload Support Structure (PSS). The PSS supports the science instruments and PM thermal enclosure. It is also acts as the telescope interface to the spacecraft.

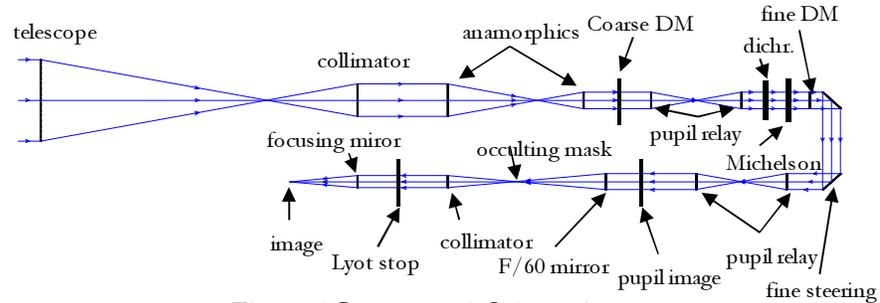

**Figure 6 Coronagraph Schematic**

**Instrument:** The starlight suppression system (Figure 6) is an expanded Lyot coronagraph with a band-limited 8th order mask. The optical design has four distinct and accessible pupil locations reserved for the Lyot mask, coarse and fine DMs, and a diffraction control optic. The coarse DM is used to control gross wavefront errors arising from gravitational release of the PM on orbit. The system operates over a bandpass of 500 – 1100 nm. Anamorphic optics provide circular beam cross section onto the coarse DM and beyond.

For FB1 only simple protected silver mirror coatings were assumed. This caused polarization leakage requiring a polarizing beamsplitter. However, compensating coating designs mitigate leakage so that the updated design replaces the polarizing beamsplitter with a dichroic, doubling system throughput. Figure 6 shows one of the two parallel arms of the coronagraph. All powered elements are used only in a collimating or focusing mode, with aberrations corrected everywhere along the optical train. The small +/- 5 arcsec field of view of the system makes it unnecessary to use more complicated optics. Filter wheels, shutters, and flip-in mirrors to feed the integral field units are not shown. The planet finding detectors are 1024 x 1024 E2V-L3 CCDs operating in photon counting charge-multiplication mode. The camera is described in detail in Goddard's CorECam instrument concept study report [12].

For exoplanet characterization, NASA Goddard designed CorSpec, an instrument consisting of four integral field spectrographs (IFS), each covering a spectral band $\Delta\lambda/\lambda \sim 22\%$ wide, and together covering the full spectral range of TPF-C. Each IFS has a $134 \times 134$ microlens array to obtain an R~70 spectrum of each Nyquist-sampled image element in the coronagraphic field, and each uses the E2V-L3 CCD to record the ~18,000 spectra. The IFS are fed by flip-in mirrors near the end of the coronagraph optical train [3].

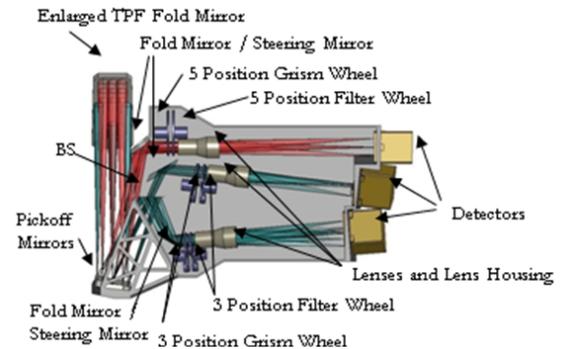

**Figure 7 Wide Field Camera (Mag30+)**

The wide field camera (Figure 7) consists of three instruments. Two cameras, one visible (425-850 nm) and one NIR (850-1700 nm) constitute a wedge-shape on the sky extending from 2-10 arcmin from the optical axis and 6 arcminutes wide. These cameras take advantage of the different orientations about the line of sight of the coronagraph to fill in a circular area 70 sq. arcmin surrounding the TPF-C target stars. A third camera works in the visible with a 4x4 arcminute field of view. These cameras provide critical context information on coronagraph observations, helping optimize overall science operations.

## 3.2 FB1 PERFORMANCE ASSESSMENT

Flowdown of the science requirements through the system design led to the top level contrast requirements shown in Table 6. The following sources of scatter are the major contributors to the detailed FB1 error budget: Static errors (assumed to be caused during assembly and fabrication); Dynamic errors driven by thermal and jitter stability during an observation including wavefront error (WFE) aberration and beam walk; and Scattered light. The error budgeting tool was later generalized and is now being used to evaluate alternative options at 2-3 $\lambda$/D. The derived engineering requirements are shown in Figure 8.

Our work in this area has led to four important lessons learned: 1) sequential wave front controllers are preferred because they relax both amplitude uniformity and surface flatness requirements; 2) Uncontrollable high-spatial frequencies are at an acceptable level using existing technology; 3) Transmissive masks placed near the image plane ahead of the coronagraph field stop have challenging surface power spectral density requirements; and 4) working at 2 or 3 $\lambda$/D is much harder than 4 $\lambda$/D resulting in several system architecture challenges, especially system stability.

Meeting these requirements has been shown to be feasible, by combination of laboratory testbed demonstrations and or by detailed system model analyses of the FB1 concept. The major result of our FB1 modeling work is that the environmental perturbations during operation appear to be controllable sufficiently— both thermally and dynamically— to ensure that the image plane contrast remains stable to the required levels. The current sunshade thermally isolates the telescope and payload adequately. Passive vibration isolation is effective, but provides less margin than an active vibration system which more robustly isolates the payload from reaction wheel vibrations (Figure 9). Vibrations from mechanisms in the instruments and starlight suppression system have yet to be explicitly addressed, but selective use of damping and observation operations reduces those risks. The FB1 launch loads impart significant reactions at the PM mounting points, and are an area of future mirror mount engineering development.

Commercial thermal and dynamic analysis software tools have limitations that FB1 design studies helped make more apparent. For longer term production use, better integrated modeling tools are being developed to improve analysis cycle time and inter-operability between the multi-physics analyses (thermal, structural, dynamics, controls and optics). Planned testbeds will validate model accuracy and predictability.

**Table 6 TPF-C Top Level Contrast Requirements for 4$\lambda$/D**

|  | Contrast | Comment |
|---|---|---|
| Static Error | 6.00E-11 | Coherent Terms |
| Contrast Stability | 2.00E-11 | Thermal + Jitter |
| Instrument Straylight | 1.50E-11 | Incoherent Terms |

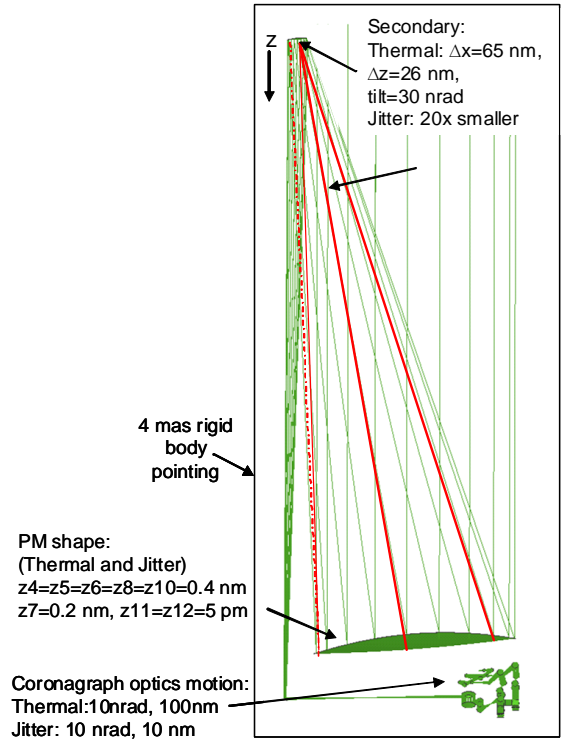

Secondary:
Thermal: $\Delta x$=65 nm, $\Delta z$=26 nm, tilt=30 nrad
Jitter: 20x smaller

4 mas rigid body pointing

PM shape:
(Thermal and Jitter)
z4=z5=z6=z8=z10=0.4 nm
z7=0.2 nm, z11=z12=5 pm

Coronagraph optics motion:
Thermal:10nrad, 100nm
Jitter: 10 nrad, 10 nm

**Figure 8    Observatory Stability Requirements**

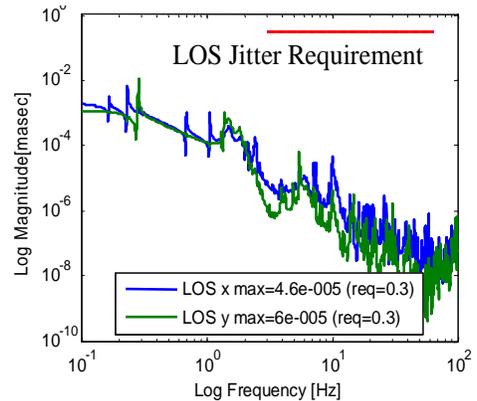

LOS Jitter Requirement

LOS x max=4.6e-005 (req=0.3)
LOS y max=6e-005 (req=0.3)

**Figure 9    Line of Sight Response with Disturbance Free Payload (DFP) Active Isolation (Lockheed Martin).**

### 3.3 VERIFICATION APPROACH

TPF-C's large size and extreme stability requirements make it impractical to fully verify system level high-contrast performance with an end-to-end test. In particular, addressing gravity effects during ground testing to the requisite precision, most importantly on the primary mirror, would certainly prove challenging, further complicated by the required thermal control and stability. Likewise, a full scale test of the thermal stability achieved by the large V-groove sunshield would require major new facility resources and would be more complicated than the flight system itself because of gravity effects on its thin tensioned membranes.

Performance verification of TPF-C will instead follow a path from subscale testbeds and test articles that validate the opto/mechanical/thermal/controls model analyses and error budget sensitivities, to subsystem tests used to correlate models in turn used to verify by analysis only the full-scale observatory end-to-end performance.

The heart of the instrument, the starlight suppression system (SSS), will be test verified at full-scale to show margins on required flight performance levels. Just as has been done with HCIT, the SSS will be illuminated with a point source representing the telescope focus, and will form a dark hole to the required $6 \times 10^{-11}$ contrast level in broadband light.

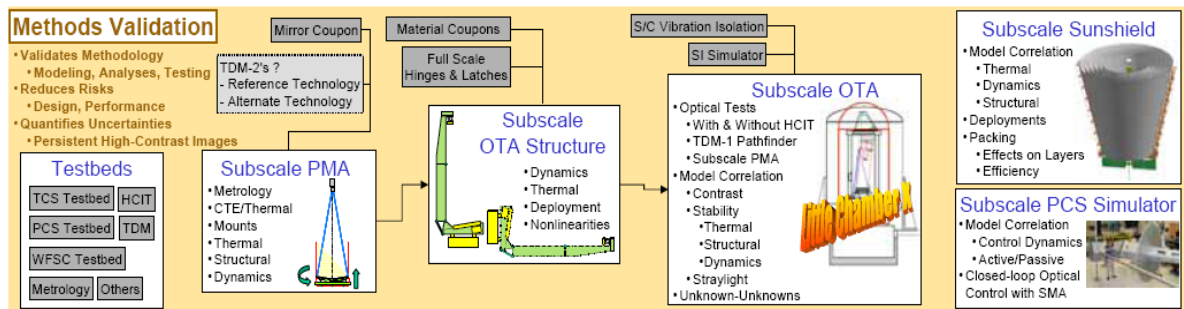

**Figure 10 TPF-C Verification Approach through subscale testbed demonstrations and model validation.**

The PM thermal control system and the primary mirror stability itself will be tested on a subscale model in a cryo-vacuum chamber such as now being upgraded at JSC. The planned test includes a center-of-curvature wavefront measurement system and high-precision metrology to monitor its position relative to the mirror.

A subscale sunshield will also be tested in a solar simulator to demonstrate its ability to reject solar energy, to maintain uniform temperature at the inside layer, and to validate thermal model predictions. It is critical that this test also includes a representative deployment mechanism because it may have a measurable impact on the performance.

Other tests include a subscale optical telescope assembly (OTA) dynamics test to demonstrate jitter stability and to evaluate nonlinearities in hinges and locking mechanism, and a subscale pointing control system simulator. These are diagrammed in the Figure 10. The schedule for these testbed activities is detailed in Figure 16. Achieving the required thermal stability is considered one of TPF-C's tall poles and is further detailed in section 4.

### 3.4 DESIGN CYCLE #2

Following the FB1 design cycle, a preliminary study demonstrated the feasibility of launching a 3.8m circular off-axis TPF-C observatory without any telescope deployments on an Atlas V (Figure 11), but further mission funding was curtailed by NASA in the early stages. The instrument is an aggressive coronagraph operating at $2.5\lambda/D$ (e.g., PIAA or Vector Vortex

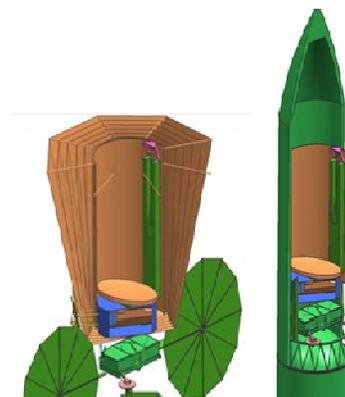

**Figure 11 TPF-C with a 3.8m telescope design**

coronagraphs) for which on-going technology development continues with very promising results.

Such an approach could provide about the same exoplanet characterizing capability as the much larger FB1 operating at $4\lambda/D$ (Figure 2), while possibly maintaining a robust general astrophysics program. A 4m-class TPF-C would decrease the cost of the mission because of its smaller telescope, no precision deployment mechanisms and a simplified integration and test program. However this comes at the expense of much tighter stability requirements. Further study is required.

## 4. KEY TECHNOLOGIES, DEVELOPMENT PLAN & PROGRESS TO DATE

Starlight suppression, precision system modeling and large space optics are the drivers of the TPF-C technology efforts. To mitigate this technical risk, TPF-C has developed a detailed technology plan [5], approved by NASA HQ, which lays out the scope, depth and inter-relatedness of activities that will enable the project to demonstrate TRL 6 by the end of Phase B.

The TPF-C technology program has established a controlled demonstration process for Technology Milestone (TM) definition and certification, applicable to any coronagraph system. Three major activities must be implemented in an integrated manner to retire technical risk by the end of formulation: 1) laboratory demonstrations, 2) validated testbed error budgets with sensitivities and 3) scaling of the error budget to flight with allocations based on the technology achievements. Underlying these TMs is model development and validation, carried out in concert with the hardware demonstrations. The details of the experiments and success criteria required to fulfill each milestone are documented in TM White Papers prior to testing, e.g. [13].

TPF-C has already completed its first two TMs, demonstrating Earth-level contrast $<1\times10^{-9}$ in monochromatic and 20% broadband light at $785\pm10$ nm within an inner region of $4$-$5\lambda/D$ and an outer region of $4$-$10$ $\lambda/D$. (Figure 12) [8,9]. The 3rd and final TM originally planned prior to TPF-C Phase A start, is underway. This is a validation of the coronagraph models and error budget sensitivities, with scaling to a flight system to demonstrate feasibility.

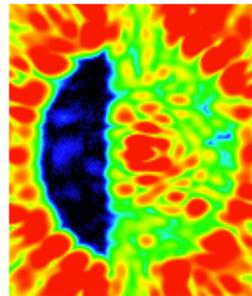

Contrast averaged across five multi-wavelength EFC iterations over a 5 hour period:

Inner 4-5 lambda/D box:
C = 5.2 e-10

Outer 4-10 lambda/D box:
C = 7.5e-10

**Figure 12 Contrast in 760-840 nm (10% bandwidth) in HCIT (Moody, et al., 2008) [7]**

### 4.1 STARLIGHT SUPPRESSION

Demonstration of the coronagraph system to reach contrast levels of $10^{-9}$ to $10^{-10}$ in broadband light over a 20% wavelength band is required for Earth imaging. The starlight suppression system consists of the coronagraph optics, deformable mirrors and wavefront control algorithms. Coronagraph model validation is also required as an indication that the physics of starlight suppression and errors driving performance including diffraction and polarization are well understood and can be extrapolated to a flight mission configuration. Contrasts of $10^{-9}$ to $10^{-10}$ in broadband light have been regularly achieved in the JPL High Contrast Imaging Testbed (HCIT) with a BL coronagraph under the TPF-C TM program. Several candidate technologies are being explored to demonstrate the feasibility of manufacturing to required tolerances: Band-limited Lyot (BL) including metallic and hybrid dielectric concepts, Shaped Pupil (SP), Visible Nuller (VN), Vector Vortex (VV) and Phase Induced Amplitude Apodization (PIAA). Of these options the metallic BL is the most mature with material options for $3\lambda/D$; SP is equally mature but suffers from low throughput; VN has comparable performance to BL but with a lower search space, VV and PIAA are emerging concepts that operate at $2\lambda/D$ with high throughput. All need further development. The TPF-C program proposes to continue exploring these approaches, with special interest given to the higher efficiency options. A down-select is expected within the first 2 years identifying the best option for a TPF-C mission.

Overall, the starlight suppression technology is estimated to be currently between TRL 3 and 5 depending on the approach, with BL being the most mature.

**Coronagraph Optics:** Manufacturing techniques for prototype amplitude masks, phase masks and PIAA optics are in hand but none, other than the metallic BLC, have been demonstrated to requirement levels. Further development will improve the optics performance to TRL 6 levels.

**Wavefront Control**: (WFC) removes spurious starlight speckles, and creates a dark hole of sufficient contrast depth to extract the image of the target planet. This is achieved in broadband light by controlling pairs of DMs to correct phase and amplitude imperfections and propagations effects. Algorithms exist to efficiently estimate the electric field and control the DM surface shape. Overall the problem of control with a perfect estimate has solid theoretical foundations, but additional development is needed to generalize the approach to 2 DMs, improve convergence speed, reduce sensitivity to sensing errors and improve robustness against partial DM failures.

**Deformable Mirrors**: (DMs) have made great strides in the last 10 years. The most notable options are the electrostrictive DMs from Xinetics Inc., and MEMS device made by Boston Micromachines Corporation. Xinetics mirrors are used at JPL's High Contrast Imaging Testbed (HCIT,) and are currently at a higher level of technology maturity (TRL 6 for 48x48mm). However, MEMS devices are attractive for mass and cost considerations. DM technologies would benefit from additional development for larger, more robust DMs (96x96mm) and require detailed calibration for WFC.

**Infrastructure**: TPF-C technology development efforts have developed critical infrastructure that benefits the demonstration of any coronagraph and has been made available to the broader community. Within the last year the TPF infrastructure has played a pivotal role in demonstrating 6 of the 7 exoplanet ASMCS mission concepts: PECO, THEIA, ACCESS, DaVINCI, EPIC, ATLAST. This infrastructure includes the HCIT, WFC systems, modeling tools for error budgets analyses, as well as mask fabrication facilities.

The **HCIT** (Figure 13) enables the exploration of starlight suppression methods and hardware in a flight-like environment within which various coronagraph concepts, WFC approaches, and control algorithms are investigated. It is uniquely capable of achieving the contrasts required for flight. The testbed is installed in a vacuum chamber and has been measured to have milli-Kelvin thermal stability and Angstrom wavefront stability. A series of increasingly mature and robust deformable mirrors have been developed, fabricated, calibrated and installed to demonstrate precise wavefront control. The testbed layout is flexible so that alternate concepts can be tried and guest investigator testing is available. The HCIT was used extensively to demonstrate the coronagraph concepts for the ASMCS studies (PECO and ACCESS).

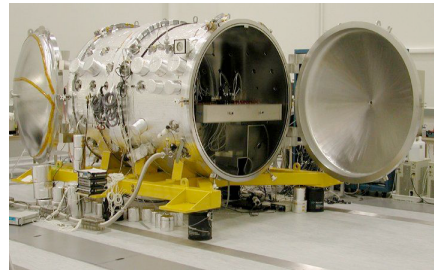

**Figure 13 HCIT chamber and bench**

### 4.2 MODELING AND VALIDATION

TPF-C will rely heavily on modeling and analyses throughout its mission lifecycle, especially since it is the primary method by which the system performance will be verified prior to launch. Thus developing, validating, and implementing models are key tasks for the project. TPF-C will make extensive use of sub-scale testbeds, not just for hardware demonstrations, but more importantly to validate models and error budget sensitivities for scaling to flight configuration. The following aspects of modeling need particular attention.

**Optical diffraction and polarization modeling with laboratory demonstration**: Models first need to adequately represent the physics to define the error budgets and flow down requirements. Above all, this involves the ability to efficiently model optical diffraction and polarization, along with their contributions from the various coronagraph components: deformable mirrors and wavefront control. Coronagraph optical modeling and validation is TM #3 in the TPF-C technology plan [5]. Model validation here involves predicting not just the end contrast, but describing the full error budget and its sensitivities to the various error contributors (e.g., alignment, wavefront errors).

To date, The BL and shaped-pupil coronagraph models are the most mature. BL models having been used extensively to guide the best-achieved contrast of 5.2 x10$^{-10}$ at 4 $\lambda$/D for 760-840nm (10% band) in natural unpolarized light in the HCIT. Other coronagraph models are less mature and need funds for demonstrations to Milestone levels.

**Integrated Modeling (IM):** Engineering development is needed to demonstrate high-fidelity models for integrated optical, structural, thermal and controls behavior, which is at the core of the end-to-end system verification prior to flight. The application of such tools includes requirements flow-down of contrast to engineering design parameters through multi-disciplinary sensitivity analyses, design optimization across several disciplines (optics, thermal, structural, jitter) and multi-disciplinary model validation during I&T. The sub-scale testbeds described in §3.3 and §4.3 will be used extensively to validate integrated modeling capabilities to the accuracy and precision required for TPF-C. Of particular importance is the ability to predict sub-nanometer optical performance to milli-Kelvin temperature stability achieved through complex multi-layer thermal control (V-groove sunshade and PM thermal enclosure). Then, especially for a deployed telescope design as is FB1, models need to represent nonlinear material and structural behaviors affecting dynamic stability. The JPL integrated modeling code CIELO, developed under partial TPF-C funding, is now parallelized such as to handle highly discretized simulations which compute temperatures to optical aberrations on the same model, well exceeding commercial thermal code size limits. CIELO will reduce numerical errors from extrapolation as well as significantly increase the analysis turn-around time compared to traditional "bucket brigade" approaches with commercial tools.

**Modeling Uncertainty:** Verification by analysis requires knowing modeling uncertainties in order to bound the predicted flight performance with respect to the requirements. A disciplined systems engineering approach will be applied to all modeling activities on TPF-C to properly capture and quantify modeling uncertainty. TPF-C error budgets will allocate modeling tolerances and reserves, just like it is traditionally done for hardware. The required model fidelity in turn defines acceptable levels of experimental errors, which themselves are verified through the error budgets of the tested articles. TPF-C will use its various testbeds to demonstrate this structured "verification by analysis" methodology and to validate modeling accuracy/predictability with its uncertainty factors.

Accurately predicting TPF-C system performance starts with the use of material data of the highest accuracy and precision. The JPL Dilatometer laboratory, co-developed by the James Webb Space Telescope (JWST) and TPF, is a state-of-the art facility which measures thermal strains from room temperature to 20°K at an accuracy of about 2 ppb. Active thermal control allows the samples to maintain a stability of 5 mK for as long as necessary, allowing the measurement of thermal relaxation and material nonlinearities. Such capability is required to measure variations in CTE distributions in Corning ULE®, nonlinear behavior of Schott Zerodur® and other materials of interest such as composites. Data from the JPL Dilatometer has been used to calibrate other CTE measurement facilities in the United States. The facility is currently mothballed for lack of funds.

### 4.3 LARGE SPACE OPTICS

The large deployable TPF-C telescope requires very high stability to enable transmission of a wavefront to the coronagraph meeting the requirements of planet finding. The technology plan has defined a series of testbeds, most of them subscale, which will demonstrate the contributions of individual components to the system level stability, reaching TRL 6 by the end of Phase B.

**Technology Demonstration Mirror** (TDM) ) is a 1.8 meter off-axis diameter mirror composed of six outer core segments and one inner hexagonal core segment, all made from Corning ULE® blanks. The TDM will demonstrate the ability to use the LTF/LTS process, developed by ITT Space Systems LLC, as described previously. The TDM will also demonstrate polishing and figuring techniques on an off-axis mirror to HST specifications. The effort will also provide methodologies to measure the performance of the mirror and to interpret the measured data for requirements verification. Owing to funding cuts, the TDM effort has been put on indefinite hold with all of the core segments completed, and the glass for the facesheets (2 prime + 1 spare) has been flowed out to size.

Coating the mirror will be a challenge because coating uniformity requirements are tight. Non-uniform coatings will cause amplitude errors that will interfere with the starlight suppression requirement. In addition, polarization effects of the candidate coatings are being studied to understand the polarization effect on starlight suppression, as well as to develop concepts for mitigation of the induced polarization of the light. For the TDM program, ITT demonstrated a capability of laying down a protected silver coating with a reflectance uniformity of 0.25% with a 3$\sigma$ uncertainty of <0.05% using witness samples held in the in a coating chamber to represent the curvature and diameter of the TDM.

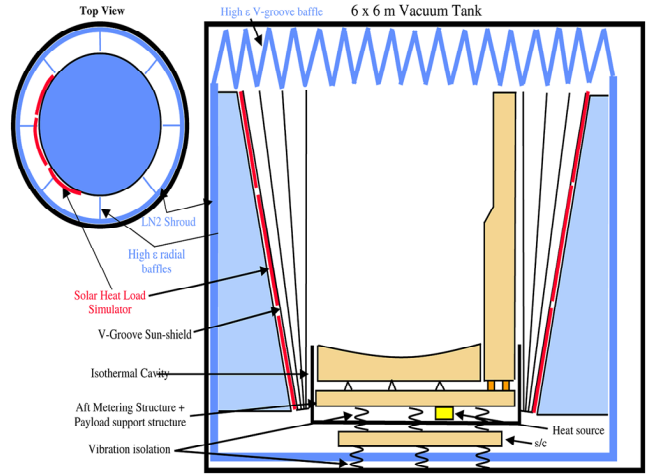

**Figure 14    Schematic of the subscale EM Sunshield and Isothermal Enclosure Testbed**

**Subscale Engineering Model (EM) Sunshield and Isothermal Enclosure**. A major source of wavefront instability is thermally induced changes from optical surfaces and in the structure linking them. The object of this testbed is to determine if, under flight-like thermal loads, the mirror temperature can be maintained stable to within the sub-milli-Kelvin limits (Figure 14). To retire this risk a quarter-scale testbed is designed incorporating the main elements of the thermal design: 1) a PM and telescope metering truss decoupled from solar radiation by a multi-layer V-groove sunshield; 2) an isothermal cavity with active thermal control which radiatively bathes the PM with a constant background flux and isothermalizes the PM and the aft metering structures (AMS) to the required precision; 3) precise thermal control of all critical conductive and radiative paths between the spacecraft and the instrument PSS.

**Sub-scale EM Primary Mirror Assembly**: This testbed is an extension of the one described above. The object is to test for wavefront stability using representative optics and changing the simulated solar illumination to approximate the dither maneuver. The mirror is of high quality ($\lambda/100$) and could be the TDM. An interferometer placed at the mirror center of curvature with provision for vibration isolation and by a laser metrology metering truss to evaluate the PM rigid body stability.

**Secondary Mirror Tower Partial Structure Testbed**: will characterize the dynamic instabilities and nonlinearities of deployable mechanisms on a full scale hinge/latch assembly with flight-like interfaces to a truncated secondary mirror (SM) tower (Figure 15) and in an environment representative of the TPF-C operating conditions. The concern is the existence of dynamic instabilities above 1 HZ which can not be corrected by the SM active hexapod and which would jeopardize the 300pm SM position stability requirement. This includes sudden and repeated energy releases of the pre-loaded mechanisms, as well as harmonic distortions of the sinusoidal jitter waveforms propagating through the nonlinear mechanisms. This testbed may not be required if a 4m-class mission design is selected for TPF-C.

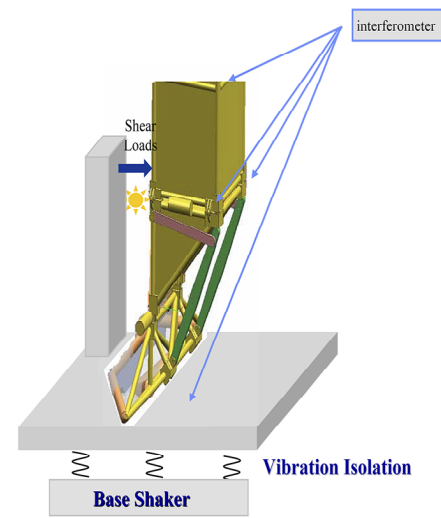

**Figure 15 Schematic of the SM Structure Testbed**

## 5.  ACTIVITY ORGANIZATION, PARTNERSHIPS AND CURRENT STATUS

**Activity Organization and Partnerships.** The TPF-C project managed by JPL, under the former Navigator Program, established strong partnerships with the exoplanet community which were severely curtailed after NASA put the project on indefinite hold.

The FB1 mission study described in this paper, was led and performed by JPL with contributions from NASA Goddard Space Flight Center (GSFC) for the telescope sub-system, from Lockheed Martin (LM) for vibration control and jitter analysis, from ITT Space Division, LLC (ITT) for primary mirror design, from Northrop Grumman Space Technologies (NGST) for the sunshield deployment design, and from Ball Aerospace for general astrophysics instrument design.

Under the TPF-C technology program many contracts were issued to university and industry collaborators for development of coronagraphic approaches and mirror technologies: Princeton University (shaped pupil masks, WFC, laboratory development), University of Hawaii / NOAO (PIAA coronagraph and mirrors), University of California Berkeley (binary masks), Harvard Smithsonian Astronomical Observatories (pupil mapping theoretical equations), University of Florida (mask fabrication), Space Telescope Science Institute (design reference mission models), ITT (TDM), Xinetics (deformable mirrors), Boston Micromachines (MEMs deformable mirrors).

In 2005, NASA also funded the TPF-C Instrument Concept Studies (ICS) as potential options for the FB1 design. The selected concepts included alternative coronagraph designs such as the VN interferometer (JPL/GSFC) and pupil remapping with anti-halo apodization (University of Arizona). They also funded a candidate primary planet detector (GSFC), a spectrometer (GSFC) and a general astrophysics wide field camera (STScI).

The STDT, lead by James Kasting from Penn State and co-chaired by TPF project scientist Wesley Traub at JPL, was composed of 24 scientists and researchers from across the nation. In 2005, NASA chartered the STDT to work with the TPF-C project to deliver 1) a mature Science Requirements Document, 2) a narrative on TPF-C's potential for general astrophysics observations, 3) a Design Reference Mission, 4) an assessment of design concepts and observational scenarios, 5) recommendations on technology development needs, 6) assistance in communicating with the astronomical community, and 7) a report summarizing their work delivered to NASA [3]. Drastic funding cuts prevented the completion of other chartered tasks: recommendations on an end-to-end science plan and on the second TPF-C design iteration.

**Current Status.** While funding for any TPF-C mission related activities has been cut, the technology program has received continued support under the newly formed Exoplanet Exploration Program (ExEP) managed by JPL for NASA, albeit at a substantially lower rate. The on-going ExEP coronagraph technology program (ExEP-C) is pursuing the development of coronagraphic technologies for starlight suppression, modeling, mask fabrication and deformable mirrors. In 2007, the activities were extended to the development of large, but efficient, optical diffraction models for external occulter tolerancing and to fabrication of sub-scale external occulter masks.

The program has successfully completed its 2 first technology milestones demonstrating 5.2x $10^{-10}$ at 4 $\lambda/D$ for 760-840nm (10% band) in natural unpolarized light using a BL, thus verifying the fundamental physics and establishing its feasibility [8, 9]. Work is on-going on TM#3 to validate optical models, error budgets and their sensitivities. Completion of TM# 3 was originally planned as the TPF-C technology gate for entry into Phase A.

More recently ExEP has provided technical infrastructure support to 6 of the 7 ASMCS coronagraph exoplanet studies as discussed above.

TPF-C intends to continue its long standing partnership with the greater exoplanet community. Under the ExEP Coronagraph Technology program, the project plans to solicit contributions through competed technology proposals for the demonstration and downselect of the starlight suppression approach as well as for contributions to the large technology testbeds. Over the mission lifecycle, NASA JPL will maintain overall management leadership of TPF-C, with competed solicitations for instruments, observatory implementation and science investigations.

## 6. MISSION SCHEDULE

Figure 16 shows the 12-year TPF-C mission schedule from pre-Phase A through launch, broken into the 4 key elements of the project: starlight suppression, telescope, thermal control and System I&T. Standard spacecraft activities are omitted from the schedule for clarity, but will be planned accordingly.

The testbeds defined in §4 are planned to begin during pre-Phase A and continued through the end of Phase B to TRL 6. These testbeds will later become part of the flight system verification process described in §3.3.

The choice of the observatory architecture size will be decided in Phase A, as it is largely dependent on the down-selected coronagraph option identified at the end of pre-Phase A. The predominant trade will be on the coronagraph IWA and its impact on telescope size and stability as well as on exoplanet science capability. Note that the development and maturation of the coronagraph technology is largely independent of the final observatory size.

The critical path for TPF-C is the procurement of raw materials needed for PM fabrication, as it is for JWST. The glass material of choice is currently Corning ULE®, selected for its low thermal expansion coefficient and stability at room temperature. Glass procurement needs to begin as early as Phase B start, right after the subscale PM technology prototype has been demonstrated to TRL 6.

## 7. TPF-C FB1 MISSION COSTS

**FB1 Mission Costs:** TPF-C is a flagship mission comparable in cost to JWST. For this mission scale, current parametric cost models can only be indicators of cost magnitude as it exceeds their domain of applicability with high uncertainty, since they are based on historical data for smaller missions. In order to estimate the costs more accurately a bottoms-up costing exercise of the FB design is necessary. Unfortunately, the program was terminated before this could be done. However, JWST provides a suitable benchmark for TPF-C FB1, having nearly the same primary mirror collecting area and for which we can trade the complexity of an actuated, segmented, cryo-figured, cryo-cooled, cryo- tested, primary mirror mounted on an ultra-stable composite backplane with that of a monolithic stiff room-temperature off-axis mirror fused from independent segment cores. The quality of the TPF-C primary mirror does not exceed that of the HST, but it must be kept extremely stable. All other elements between JWST and TPF-C are of comparable magnitude: sunshield, integration and test, instrument module, and ground segment. The procurement of the PM material and its fabrication will be a long lead item for TPF-C, as it has been for JWST. Hence the funding profiles will essentially be identical.

**Technology Costs:** Since 2003, the TPF-C technology program has benefited from approximately $36M NASA investment, of which nearly half has been distributed to the community to develop and test novel coronagraph concepts, deformable mirrors, and begin demonstrating precision primary mirror optics fabrication (TDM). Infrastructure support has also been provided to demonstrate the ASMCS concepts. This overall investment is summarized in Table 7.

During the first 5 years, coronagraphic technology development and demonstration costs are estimated to be around $200M to bring all components to TRL 6, as shown on the schedule in Figure 16. Specifically, in addition to the funds spent to date, approximately $50M should be allocated for in depth demonstration of the most promising starlight suppression approach to TRL 6, with an option downselect within the first 2 years. Currently the PIAA and optical vortex are good candidates for a concept at smaller inner working angles, and the BL is a more mature option imposing relaxed stability on the design. The remainder of the costs will be allocated to the large testbeds, especially the TDM and the subscale EM Sunshield/Primary mirror testbed.

**Figure 16    TPF-C Flight Baseline 1 Mission Schedule, Including Technology Development**

Work on these testbeds can begin in parallel with the starlight suppression demonstrations, as they are largely independent of coronagraph type. The   deployment testbed may not be required if a smaller 4m-class TPF-C design is selected since it has no telescope deployment.

**Cost and Architecture Considerations:** A 4m-class circular aperture concept for TPF-C would provide cost savings with possibly limited impact to the exoplanet science if a coronagraph concept at $2.5\lambda/D$ proves to be feasible (Figure 2). First since the telescope cost scales as roughly $D^{2.5}$, we expect a factor of about 5 in reduction in telescope cost alone. A 4m-class concept can be launched without deployment, thus reducing the number of mechanisms, their associated risk and required technology development. While maintaining stability is tighter at smaller IWAs, the environment will be more stable since the telescope will not have to roll about its axis to fill a circular image from the elliptical PM. Furthermore there will be no deployments in the telescope to create dynamic nonlinearities. Additional studies are required to determine the feasibility and merits of a 4m-class concept compared to FB1.

**Table 7 TPF Coronagraph Technology Investments (2003-09)**

| | $K |
|---|---|
| **TPF-C TECHNOLOGY DEVELOPMENT &** | |
| HCIT Infrastructure & Tests | $    13,800 |
| Optics Modeling | $         920 |
| Lyot & Shaped Pupil Mask Design & Fabrication | $      2,400 |
| Integrated Modeling Tools | $      1,000 |
| Material Characterization | $           80 |
| Structural Stability | $           70 |
| **TOTAL** | **$    18,300** |
| **CORONAGRAPH COMMUNITY SUPPORT** | |
| Tech Demonstration Mirror | $    10,000 |
| Xinetics DM | $      2,500 |
| MEMs DMs | $         560 |
| PIAA Coronagraph | $         880 |
| Shaped Pupil Coronagraph | $         920 |
| Visible Nuller Coronagraph | $      2,200 |
| Binary Mask Coronagraph | $         290 |
| Mask Fabrication | $         170 |
| DRM Models | $         240 |
| **TOTAL** | **$    17,800** |
| **TOTAL TPF-C CORONAGRAPH** | |
| | **$    36,100** |

## 8. CONCLUSIONS

TPF-C is scoped to perform the observations that the field of exoplanet research desires - high SNR spectra of a significant number of Earthlike planets around nearby stars, without compromises or questionable assumptions. When it eventually flies, TPF-C will be one of the most scientifically exciting missions ever launched by NASA. A positive indication of extraterrestrial life, or even the detection of a habitable planet similar to Earth, would alter the way in which humans look at themselves and at the universe.

The required technologies are well on their way to completion, and the FB-1 design shows that they can be transformed into a capable science instrument. Coronagraphic techniques required to achieve $10^{-10}$ starlight suppression at close distances to the star, can likely be developed in the near-term given a modest amount of funding.

What is needed? Astro2010 should endorse a steady effort to complete the TPF-C technology program and a resumption of a mission design that brings online a powerful tool for astronomers to search the nearby universe for planets like ours.

## 10. ACKNOWLEDGEMENTS

The work described in this report was performed at the Jet Propulsion Laboratory, California Institute of Technology, under a contract with the National Aeronautics and Space Administration.
© 2009 California Institute of Technology. Government sponsorship acknowledged.